\documentclass[12]{article}
\usepackage{amsmath}
\usepackage{amsfonts}
\input psfig.sty
\def\fig{.}

\def\Iden{\mbox{$\bf 1\ $}}
\def\n{\noindent}

\def\H {\mathcal{H}}

\begin{document}

\baselineskip .5cm

\author{Navin Khaneja,\thanks{Division of Applied Sciences, Harvard 
University, Cambridge, MA 02138. Email:navin@hrl.harvard.edu. This 
work was funded by DARPA-Stanford grant F49620-01-1-0556.}\ \ \ 
Steffen J. Glaser \thanks{Institute
of Organic Chemistry and Biochemistry II, Technische Universit\"at M\"unchen,
85747 Garching, Germany. This work was funded by the Fonds der 
Chemischen Industrie and the Deutsche Forschungsgemeinschaft under 
grant Gl 203/4-1.}}

\title{{\bf Efficient Transfer of Coherence through Ising Spin Chains}}

\maketitle
\begin{center}
{\bf Abstract}
\end{center}
\n Experiments in coherent spectroscopy correspond to control of quantum mechanical ensembles
guiding them from initial to final target states by unitary 
transformations. The control
inputs (pulse sequences) that accomplish these unitary transformations 
should take as little time as possible so as to minimize the effects of relaxation and to optimize the
sensitivity of the experiments. Here, we present a novel approach for efficient control of dynamics in spin chains 
of arbitrary
length.  The approach relies on creating certain three spin encoded states, which can be efficiently transferred through a spin chain. The methods presented are expected to find applications in
control of spin dynamics in coherent spectroscopy and quantum 
information processing.

\begin{center}
\section{Introduction}\end{center}
According to the postulates of quantum mechanics, the evolution of 
the state of a
closed quantum system is unitary and is governed by the 
time-dependent Schr{\"o}dinger
equation. This evolution can be controlled by systematically changing
the Hamiltonian of the system. The control of quantum systems has important
applications in physics and chemistry
$[1$-$4]$.  In
particular, the ability to steer the state of a  quantum system (or 
of an ensemble of
quantum  systems) from a given initial state to a desired target 
state forms  the basis
of spectroscopic techniques such as nuclear magnetic resonance (NMR) 
and electron spin
resonance (ESR) spectroscopy \cite{Ernst, electron}, laser coherent
control
\cite{optics} and quantum computing
\cite{QC1, QC2}. Achieving a desired unitary evolution in a quantum system in minimum time is an important practical problem in 
coherent spectroscopy. Developing short pulse sequences (control laws) which
produce a desired unitary evolution has been a major thrust in NMR
spectroscopy \cite{Ernst}. For example, in the NMR spectroscopy of 
proteins, the
transfer of  coherence along spin chains is an
essential step in a large number of key experiments.

A spectroscopist has at his disposal a limited set of control 
Hamiltonians $\{ \H_j \}$
(produced by external electromagnetic fields) that can be turned on 
and off to modify
the net Hamiltonian of the system. There is a natural coupling 
(interaction) between the
spins and in the absence of any external control Hamiltonians, the 
state of spin system
evolves under this interaction or coupling Hamiltonian $\H_c$. The 
task of the pulse
designer is to find the right sequence of external pulses 
interspersed with evolution of the 
system under the coupling Hamiltonian $\H_c$ for different time 
periods, in order to
create a net evolution or unitary transformation that transforms the 
state of the system
from some initial to a desired final state in minimum possible time.

\n Even for two coupled spins 1/2, the time-optimal
transfer of polarization or of coherence is not trivial
\cite{bound, time.opt}. Numerous approaches have been proposed and
are currently used
\cite{Cavanagh} to transfer
polarization or coherence through chains of coupled spins.
Examples are the design of radio-frequency (RF)
pulse trains that create an effective Hamiltonian \cite{Ernst, Advances}, which makes it possible to
propagate spin waves
in such chains \cite{XYa, XYb, spinwave1, spinwave2, Brueschweiler}.
In order to achieve the maximum possible
transfer amplitude, many other approaches, that
rely either on a series of
selective transfer steps between adjacent spins or on 
concatenations of two such selective transfer steps \cite{Cavanagh, Advances} have been developed.

Here, we consider a novel approach to control the transfer of coherence
in spin chains of arbitrary length. The approach relies on the
creation of a three spin encoded state and efficient propagation
of this encoded state through the spin chain.
Our method is based on variational
ideas as captured by the theory of optimal control \cite{time.opt}.
In the present context of control of nuclear 
spin ensembles,  we
are interested in finding a sequence of RF pulses that will efficiently transfer any state (known or unknown) of a given spin in a spin chain to a desired target spin.  Compared to conventional experiments, this new approach makes 
it possible to
speed up the transfer rate by up to  a factor of three, which suggests
applications in NMR spectroscopy and 
experimental quantum computation.

\n Without loss of generality, here we consider the problem of transferring the coherence of a spin at one 
end of a spin chain
(label spin 1) to a spin at the opposite end of the chain (label spin 
n) in a spin
ensemble. Consider an initial density operator $\rho_0$ representing coherence or polarization on the first spin, which has the general form $\rho_0 = \frac{1}{2} \Iden + a_1 I_{1x} + a_2 I_{1y} + a_3 I_{1z}$. The goal is to transfer this density operator to the operator $\frac{1}{2} \Iden + a_1
I_{nx} + a_2 I_{ny} + a_3 I_{nz}$. Note that it suffices to find a unitary
transformation
$U$ that transfers $I_{1x} \rightarrow I_{nx}$ and $I_{1y} 
\rightarrow I_{ny}$. The same
unitary transformation will also transfer $[I_{1x}, I_{1y}] 
\rightarrow [I_{nx},
I_{ny}]$, i.e. $I_{1z} \rightarrow I_{nz}$. Therefore, the transfer of 
the coherent state
of spin $1$ to spin $n$ is equivalent to the transfer of the 
non-Hermitian operator
$I^-_1 = I_{1x} - i I_{1y}$ to $I^-_n = I_{nx} - i I_{ny}$. The 
transfer between such
non-Hermitian operators arises naturally in coherent spectroscopy of ensembles and 
constitutes a
fundamental step in  multidimensional NMR spectroscopy of biological 
macro-molecules
\cite{bound}. We emphasize again that any unitary transformation that transfers $I^-_1$ to $I^-_n$, implies 
that any state of the form $\rho_0 = \frac{1}{2} \Iden + a_1 I_{1x} + a_2 I_{1y} + a_3 I_{1z}$ is transferred to spin $n$.

\n Besides applications in spectroscopy,
finding optimal methods to control
the dynamics of coupled spin networks is of fundamental importance for the
practical implementation of quantum information processing. In recent years,
many innovative proposals have come
out to harness the dynamics of spins in the liquid \cite{QC1, QC2} 
and solid state
\cite{kane, yamamoto} for the purpose of information processing. Like 
many coherence
transfer experiments in multidimensional NMR spectroscopy, these NMR 
quantum computing
architectures rely on elaborate sequences of radio frequency (RF) 
pulses  for realizing
desired effective Hamiltonians. Recent proposals by Yamamoto
\cite{yamamoto} use a chain of nuclear spins $\frac{1}{2}$ in the solid state
for purpose of computing. A major challenge in such
architectures, which is also found in various other quantum information
devices, is finding efficient ways of making qubits interact if they 
are not directly
coupled. A prototype example of this problem is finding efficient
ways to generate unitary transformations which exchange the states of 
spins on the two
opposite ends of a spin chain. Pulse sequences presented in this 
paper can be used to
accomplish such operations efficiently.

\begin{center}{\bf Control of Spin Chain Dynamics} \end{center}
Consider a linear chain of $n$ weakly interacting spin $\frac{1}{2}$
particles placed in a static external 
magnetic field in the $z$
direction  and with Ising type couplings between next neighbors \cite{Ising1925,Caspers}.
In a
suitably chosen (multiple) rotating frame which rotates with each 
spin at its
resonant frequency, the Hamiltonian that governs the free evolution of the spin
system is given by the coupling Hamiltonian
$$
\H_c = 2
\pi
\sum_{k=2}^{n} J_{k-1,k}\ I_{(k-1)z}I_{kz},$$ where $J_{k-1,k}$ is the coupling
constant between spin
$k-1$ and
$k$. If the resonance frequencies of the spins are well separated, 
spin $k$ can be
selectively excited (addressed) by an appropriate choice of the 
amplitude and phase of
the RF field at its resonant frequency. The goal of the pulse 
designer is to make
appropriate choice of the control variables comprising of the 
frequency, amplitude and
phase of the external RF field to effect a net unitary evolution 
$U(t)$ which efficiently transfers
the initial operator $A=I_{1}^{-}$ to $B=I_{n}^{-}$. For
simplicity but without loss of generality (vide infra), we assume that
all
coupling constants in the spin chain are equal, i.e. $J_{k-1,k}=J$ 
for $1 < k \leq
n$.

\begin{figure}[t]
\centerline{\psfig{file= \fig/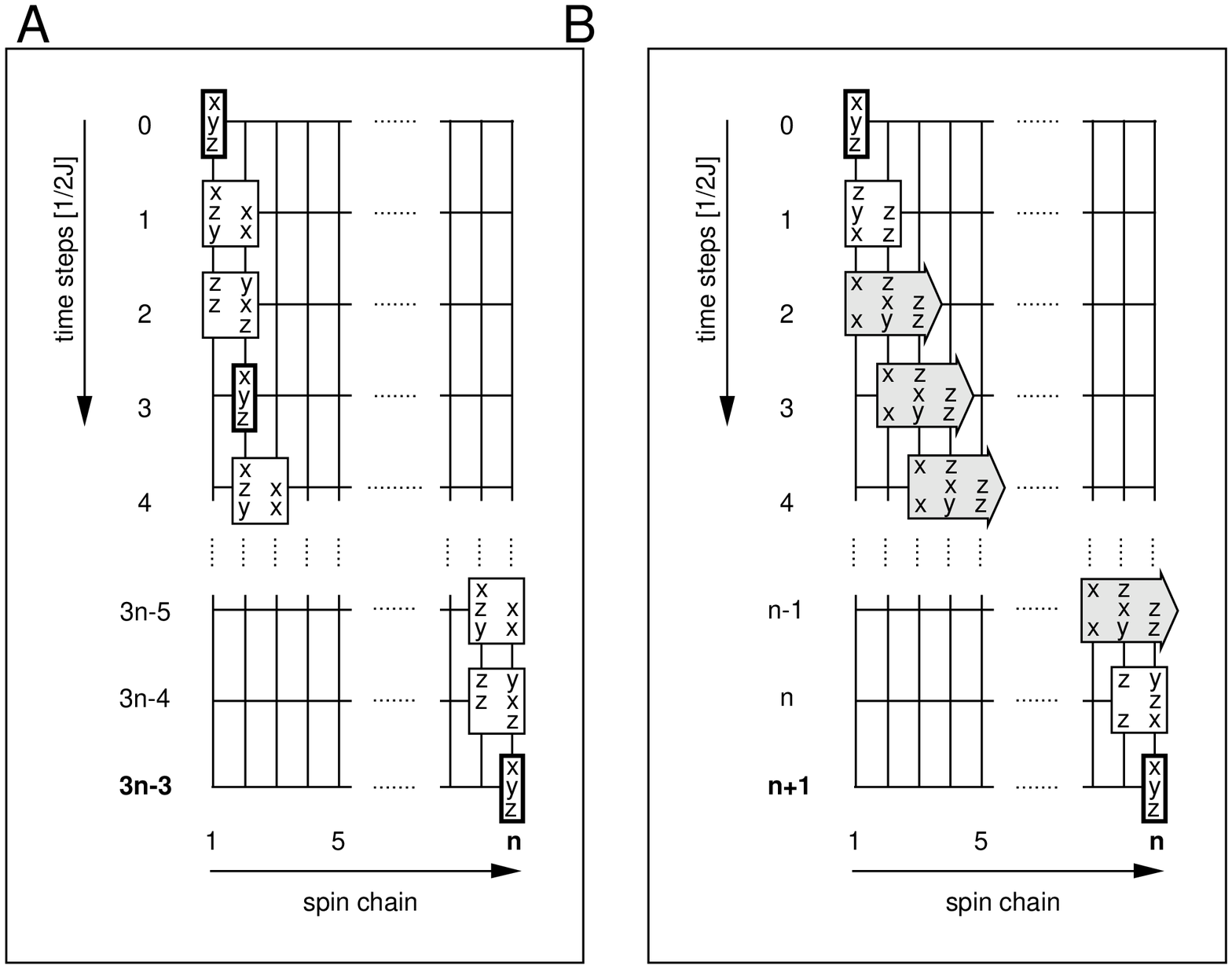 ,width=6in}}
\caption[coset]{Panel A
shows the flow of coherence and polarization in a spin chain under a 
sequence of selective isotropic
mixing periods \cite{Advances, Isotropic1, Isotropic2}, (sequential swap operations)
each of which can be decomposed in three steps of duration 
$\Delta=(2J)^{-1}$ with effective Hamiltonians
$2 \pi J I_{(k-1)x}I_{kx}$, $ 2 \pi J I_{(k-1)y}I_{ky}$, and $2 \pi J 
I_{(k-1)z}I_{kz}$, respectively.
After the first step the initial operators
$I_{1x},\ I_{1y},\ I_{1z}$ are transferred to $I_{1x},\ 2 
I_{1z}I_{2x},\ 2 I_{1y}I_{2x}$ respectively. Coherence
is transferred in a sequential manner where the state of the spin $k$ 
is transferred to  spin
$k+1$ in
$\frac{3J}{2}$ units  of time. The total transfer takes 
$\frac{3(n-1)}{2J}$ units of time. Panel B
shows the flow of coherence and polarization under 
the proposed pulse sequence based on effective
soliton operators (indicated by grey arrows). Here, a localized  spin 
wave is created which moves one step in the
spin chain in every
$\frac{1}{2J}$ seconds. The total transfer time for the proposed 
pulse sequence is
$\frac{n+1}{2J}$. For clarity, operators such as $I_{kx}$ are indicated by the 
letter $x$ at position $k$. Similarly,
bilinear (or trilinear) product operators such as $\Lambda_{kx}$ (or 
$\Lambda_{kz}$) are indicated only by the
axis labels $x$, $y$, or $z$ at the corresponding spin position, 
omitting prefactors of 2 (or of 4) and
possible algebraic signs.
\label{fig:coherence.flow}}
\end{figure}

A straight forward way of transferring the operator $I^-_1$ to $I_n^-$ is to perform sequential transfers, whereby $I_k^-$ is transferred to $I_{k+1}^-$ \cite{Isotropic1, Isotropic2} . Each of these sequential steps takes $\frac{3}{2J}$ units of time, resulting in a total time of $\frac{3(n-1)}{2J}$ (see Fig.\ref{fig:coherence.flow}). However this is far from optimal. It can be shown \cite{time.opt} that given any initial operator $A_k$ that represents
a state involving spin $k$ and spins with label less than 
$k$, the minimum time required to advance this operator one step up the spin chain is $\frac{1}{2J}$. This suggests that an efficient approach to transferring the state $I^-_1$ to $I_n^-$ is to prepare an encoded state $\Lambda_k^{-}$ such that it is possible to go from
$\Lambda_k^{-}$ to $\Lambda_{k+1}^{-}$ in a time of
$\frac{1}{2J}$. Furthermore, it is desirable that these encoded states be sufficiently localized
so that $\Lambda_1^{-}$ can be encoded and decoded in a short time.
We will refer to such encoded states as effective soliton operators (see end of this section).

Now we consider the three specific effective soliton operators
${\Lambda}_{kx}=2 I_{(k-2)x} I_{(k-1)z}$, $\Lambda_{ky}=2 I_{(k-1)x}
I_{kz}$, and  $\Lambda_{kz}=4 I_{(k-2)x}
I_{(k-1)y}I_{kz}$, which obey the commutation relations $\lbrack
\Lambda_{\alpha},
\Lambda_{\beta}\rbrack = i
\epsilon_{\alpha
\beta \gamma}\Lambda_{\gamma}$, where $\epsilon_{\alpha \beta 
\gamma}$ is the Levi-Civita symbol which is 1
(or $-1$) if $\{\alpha \beta
\gamma\}$ is an even (or odd) permutation of $\{x,y,z\}$ and 0 if two
or  more of the indices
$\alpha$, $\beta$,  $\gamma$ are identical.
Each individual soliton operator ${\Lambda}_{k\alpha}$ is advanced
along the spin chain by one unit if
the
propagator
\begin{equation} \label{eq:main} U_\Lambda=\ \exp\{- {\rm i}\Delta 
{\cal H}_c\}\  \exp\{- {\rm i}
{\pi \over 2}  F_y\}\end{equation}
with $F_y =I_{1y}
+ I_{2y} + ...
I_{ny}$
is applied:
$${\Lambda}_{k\alpha} {\buildrel {U_\Lambda} \over \longrightarrow}
   {\Lambda}_{(k+1)\alpha}.$$
The propagator $U_\Lambda$
can be realized by applying a non-selective 90$^\circ_y$ pulse
(with negligible duration) to all spins, followed by the
evolution of the spin system under the coupling Hamiltonian ${\cal
H}_c$ for a duration $\Delta=(2J)^{-1}$.

With the help of the soliton operators $\Lambda^-_k = {\Lambda}_{kx}
- {\rm i} {\Lambda}_{ky}$, it is possible to
transfer  $I^-_1=I_{1x}-{\rm i} I_{1y}$ efficiently to
$I^-_n=I_{nx}-{\rm i}
I_{ny}$:
$$I^-_1 \ \
{\buildrel {U_{1}} \over \longrightarrow}
{\buildrel {U_{2}} \over \longrightarrow} \ \ \ \
\Lambda^-_3 \ \ \
\underbrace{{\buildrel {U_\Lambda} \over \longrightarrow} \ \  ..... \ \
{\buildrel {U_\Lambda} \over \longrightarrow}}_{(n-3) {\rm times}}
\  \ \
\Lambda^-_{n} \ \ \ \
{\buildrel {U_{\Lambda}} \over \longrightarrow}
{\buildrel {U_{{ n+1}}} \over \longrightarrow} \ \
I^-_n. \
$$
Here, the encoding of $I^-_1$ as the soliton operator
$\Lambda^-_3$ is effected by the propagators
$$ U_1=
\exp\{- {\rm i}\Delta {\cal H}_c\}
\exp\{ {\rm i} {\pi \over 2} I_{1y}\}
\exp\{ {-\rm i} {\pi \over 2} I_{1x}\},
$$ and $$U_2=\exp\{- {\rm i}\Delta {\cal H}_c\}
\exp\{- {\rm i} {\pi \over 2} (I_{1x}+I_{2y})\}.$$ Finally, the 
decoding of the the soliton operator $\Lambda^-_{n}$ into
$I^-_n$ is achieved by the propagators $U_\Lambda$ and
$U_{n+1}=
\exp\{ {\rm i} {\pi \over 2} I_{nx}\}
\exp\{- {\rm i}\Delta {\cal H}_c\}
\exp\{{\rm i} {\pi \over 2} (I_{nx} - I_{(n-1)y})\}.
$
${U_1}$, ${U_2}$, ${U_\Lambda}$, and ${U_{n+1}}$ require a period 
$\Delta=(2J)^{-1}$ each, resulting
in the time
$$\tau_{soliton}={{n+1}\over{2J}}$$
for the complete transfer from  $I^-_1$ to  $I^-_n$.

The flow of soliton operators is summarized in Fig.2B. 
 The panel schematically traces the 
evolution of the initial operators
  $I_{1x},\ I_{1y}$, and $I_{1z}$ via the soliton operators 
$\Lambda_{kx}$, $\Lambda_{ky}$, and $\Lambda_{kz}$
in the spin chain as a function of time. The operators $\Lambda_{kx}$ , $\Lambda_{ky}$ and $\Lambda_{kz}$ represent local correlations of spin $k-1$ with its neighbors. Under the proposed pulse sequences, these correlations advance one step in the spin chain, every $\tau_{step} = \frac{1}{2J}$ units of time. Although these operators evolve to other operators under the proposed pulse sequences, if the spin system is observed stroboscopically, every $\tau_{step}$ units of time, the correlations maintain their shape and are just translated one step up in the spin chain. Hence the name effective soliton operators.

\begin{center}{\bf Efficiency of Transfer} \end{center}

\begin{figure}[t]
\centerline{\psfig{file= \fig/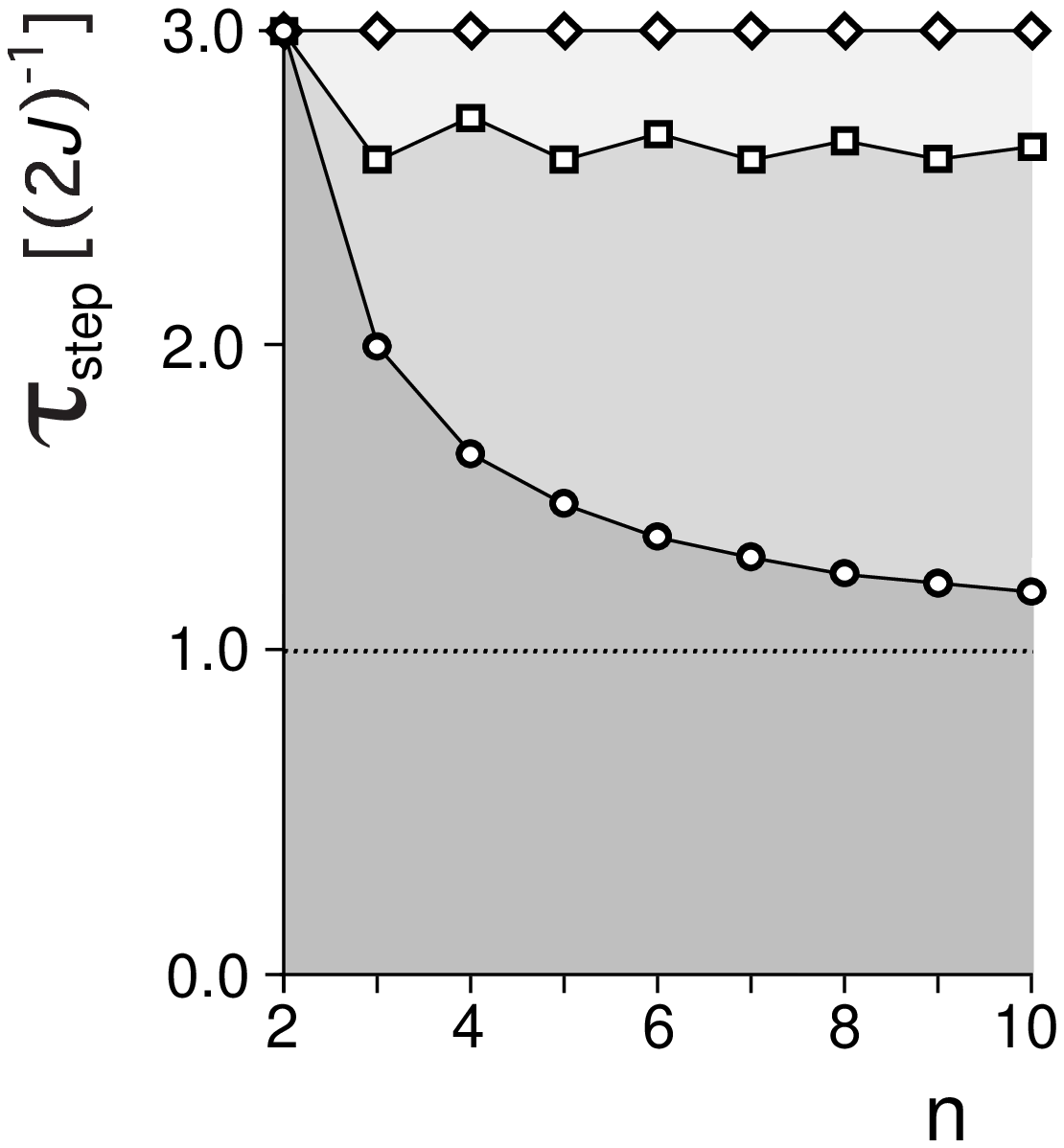 ,width=4in}}
\caption[coset]{The figure shows a comparison of $\tau_{step}=\tau/(n-1)$, the 
average time required to advance by one step in a chain of $n$ 
coupled spins for pulse sequences which effect full transfer from 
$I_1^-$ to $I_n^-$.
Diamonds:
conventional sequence of selective isotropic mixing steps between 
neighboring spins \cite{Isotropic1, Isotropic2}.
Squares:
Sequence of optimal indirect SWAP(k, k+2) operations \cite{3-spin} 
(of duration $\frac{3 \sqrt{3}}{2J}$) which are followed by a selective 
isotropic mixing steps between 
spins $(n-1)$ and $n$ if $n$ is even.
Circles: Soliton pulse sequence.
\label{fig:comparison}}
\end{figure}

\n The time $\tau_{soliton}$ taken by the proposed pulse sequence should be
compared with the transfer time for conventional pulse sequences
which transfer $I_1^{-}$ to $I_n^{-}$ \cite{Advances, Isotropic1,
Isotropic2}.
These pulse sequences require $n-1$ steps of selective isotropic transfers in
which the $j^{th}$ step transfers the operator $I_j^{-}$ to $I_{j+1}^{-}$.
In the $j^{th}$ step, only spins $j$ and $j+1$ are active and the remaining
spins in the chain are decoupled. This mode of transfer is depicted 
in panel A of
Figure \ref{fig:coherence.flow}. Each such isotropic transfer step
requires $\frac{3}{2J}$ units of time and therefore
the total time is $\frac{3(n-1)}{2J}$. In the limit of large $n$,
the proposed soliton sequences only take $\frac{1}{3}$ amount of time 
as compared
to state of the art pulse sequences. A comparison of the time taken for
the coherence transfer by the conventional sequence of selective 
isotropic pulse sequences $\tau_{conv}$ and the
proposed pulse sequences $\tau_{soliton}$ is shown in the
Figure \ref{fig:comparison} for $n\leq 10$.

The proposed pulse sequences are also compared with the
widely used
concatenated INEPT pulse sequences \cite{concatenation}, which transfer only
one component of
magnetization along a spin chain, i.e. $I_{1x} \rightarrow I_{nx}$. If all
couplings are equal to $J$, the time required for
transferring $I_{1x} \rightarrow I_{nx}$ by the
concatenated INEPT pulse sequences is $\frac{n}{2J}$. For large $n$, this is
approximately the same as the time required for the new soliton-based pulse
sequences. However, the soliton sequences transfer the complete state 
of spin $1$ to spin $n$, which may
result in appreciable gain of signal to noise ratio in spectroscopic 
applications \cite{Cavanagh}.

The proposed pulse sequences can be used to efficiently exchange the
(arbitrary) states of spins at the two ends of a spin chain with possible applications
to proposed quantum computing architectures \cite{yamamoto}.
This exchange operation between spin $1$ and $n$, in general will not preserve the state of other spins on the spin chain and hence donot represent a swap gate between spin $1$ and $n$ in the usual sense.
The proposed soliton sequences can be used to 
transfer the
state of spin 1 to spin $n$ and spin $n$ to spin $1$ simultaneously in $\frac{n+1}{2J}$ units of time. This should be compared to the approach where states of spin $1$ and $n$ are exchanged through a sequence of neighbouring swap operations \cite{thomas}. Each swap operation requires $\frac{3}{2J}$ units of time. Therefore 
the total time required to exchange states of
spin $1$ and $n$ by sequential swapping is at least 
$\frac{3(n-1)}{2J}$, which in the limit of large $n$ is three times longer than the proposed approach based on soliton operators. 

Finally, we analyze the time taken by the proposed coherence
transfer methodology when couplings are not equal. To produce the effect of
the propagator $U_{\Lambda}$ acting on $\Lambda_{kx}, \Lambda_{ky}, 
\Lambda_{kz}$ in equation \ref{eq:main}, the only terms in the coupling Hamiltonian $\H_c$ that 
are instrumental are
$$ 2 \pi (J_{k-2,k-1} I_{(k-2)z}I_{(k-1)z} + J_{k-1,k} I_{(k-1)z}I_{kz} + J_{k,k+1} I_{kz}I_{(k+1)z}). $$ If 
$J_{k, k+1}$ is the smallest of the coupling constants $J_{k-2, k-1}, J_{k-1, 
k}, J_{k, k+1}$, i.e. if $$ J_{k, k+1} = \mbox{min} (J_{k-2, k-1}, J_{k-1, 
k}, J_{k, k+1}), $$ then it will take $\tau_k = \frac{1}{2 
J_{k, k+1}}$ units of time to produce
  the propagator $$ \exp\{- {\rm i}\pi (I_{(k-2)z}I_{(k-1)z} + 
I_{(k-1)z}I_{kz} + I_{kz}I_{(k+1)z}) \}.$$ This is achieved by 
letting the coupling $J_{k,k+1}$ evolve during
$\tau_k$ while letting
$J_{k-2, k-1}$ and $J_{k-1, k}$ couplings evolve only during $\frac{J_{k-2, 
k-1}}{J_{k, k+1}}\tau_k$ and $\frac{J_{k-1, k}}{J_{k, k+1}}\tau_k$ 
respectively and decoupling these couplings for the 
remaining time. This can be achieved by
standard refocusing techniques \cite{Ernst}. Therefore the total time 
required for propagation of the soliton $\Lambda_3$ to $\Lambda_n$ is 
$$ \sum_{k=1}^{n-3}\frac{1}{2 \ \mbox{min} (J_{k, k+1}, J_{k+1, k+2}, 
J_{k+2, k+3})}.$$ Similar arguments yield that the time
  required for preparation of soliton state from the initial state is
$\frac{1}{2 J_{12}} + \frac{1}{2 \ \mbox{min} (J_{12}, J_{23})}$ and finally
  the time required to reduce the soliton state to the final state is
$\frac{1}{2 \ \mbox{min} (J_{n-2, n-1}, J_{n-1, n})} + \frac{1}{2 \ 
J_{n-1, n}}$.

The proposed methods for control of spin chain 
dynamics may have the potential to
improve the senstivity of multi-dimensional heteronuclear
triple resonance experiments, used for example for sequential resonance assignments in
protein NMR spectroscopy \cite{Cavanagh}. The proposed method of manipulating dynamics of spin chains could also reduce decoherence
effects in experimental realizations of quantum information devices
\cite{yamamoto}. Although minimizing the time required to produce desired unitary evolutions in a quantum system is expected to reduce dissipation and relaxation effects, optimizing pulse sequences by incorporating a realistic relaxation model may further improve the sensitivity of experiments in coherent spectroscopy.


\begin{thebibliography}{99}

\bibitem{Ernst}
R. R. Ernst, G. Bodenhausen, A. Wokaun, {\it Principles of Nuclear Magnetic Resonance in One and Two Dimensions}, (Clarendon Press, Oxford, 1987).

\bibitem{wiseman} H.M. Wiseman, G.J. Milburn, {\it Phys. Rev. 
Lett.} {\bf 70}, 548 (1993).

\bibitem{electron} A. Schweiger, in {\it Modern Pulsed and Continuous 
Wave Electron Spin Resonance}, M.K. Bowman, Ed. (Wiley, London, 
1990), pp 43.

\bibitem{optics}
W. S. Warren, H. Rabitz, M. Dahleh, {\it Science.} {\bf 259}, 1581 (1993).

\bibitem{QC1} D. G. Cory, A. Fahmy, T. Havel, {\it Proc. Natl. Acad. 
Sci. USA.}   {\bf 94}, 1634  (1997).

\bibitem{QC2} N. A. Gershenfeld, I. L. Chuang, {\it Science} {\bf 
275}, 350 (1997).

\bibitem{bound}
S. J. Glaser et al., {\it Science} {\bf 208}, 421 (1998).

\bibitem{time.opt} N. Khaneja, R.W. Brockett, S.J. Glaser, {\it Phys. Rev. A} {\bf 63}, 032308 (2001).

\bibitem{3-spin} N. Khaneja, S.J. Glaser, R.W. Brockett, 
{\it Phys. Rev. A} {\bf 65}, 032301 (2002).

\bibitem{Cavanagh}
J. Cavanagh, W. J. Fairbrother, A. G. Palmer, N. J. Skelton,
{\it Protein NMR Spectroscopy: Principles and Practice},
(Academic Press, San Diego, 1996. )

\bibitem{Advances}
S. J. Glaser, J. J. Quant, {\it in Advances in
Magnetic and Optical Resonance}, W. S. Warren, Ed. 
(Academic Press, San Diego, 1996), Vol. 19, pp. 59.

\bibitem{XYa}
E. H. Lieb, T. Schultz, D. C. Mattis, {\it Ann. Phys. (Paris)} {\bf 16}, 
407 (1961).

\bibitem{XYb}
H. M. Pastawski, G. Usaj, P. R. Levstein,
{\it Chem. Phys. Lett.} {\bf 261}, 329 (1996).

\bibitem{spinwave1}
R. M. White, {\it Quantum Theory of Magnetism} (Springer, Berlin, 1983).

\bibitem{spinwave2}
D. C. Mattis, {\it The Theory of Magnetism I, Statistics and Dynamics}
(Springer, Berlin, 1988).

\bibitem{Brueschweiler}
Z. L. M\'adi, B. Brutscher, T. Schulte-Herbr\"uggen, R. Br\"uschweiler, 
R. R. Ernst, {\it Chem. Phys. Lett.} {\bf 268}, 300 (1997).

\bibitem{kane} B.E. Kane, Nature {\bf 393}, 133 (1998).

\bibitem{yamamoto} F. Yamaguchi, Y. Yamamoto, {\it Appl. Phys. A} {\bf 68} (1999).

\bibitem{Ising1925}
E. Ising, {\it Z. Physik}
{\bf  31}, 253 (1925).

\bibitem{Caspers}
W. J. Caspers, {\it Spin Systems} (World Scientific, London, 1989).

\bibitem{thomas} T. Schulte-Herbr\"uggen, O.W. S\o rensen, {\it Concepts Magn. Reson.} 12 389 (2000). 

\bibitem{Isotropic1}
D. P. Weitekamp,  J. R. Garbow,  A. Pines,
{\it J. Chem. Phys.} {\bf 77}, 2870 (1982).

\bibitem{Isotropic2}
P. Caravatti,  L. Braunschweiler,  R. R. Ernst,
{\it Chem. Phys. Lett.} {\bf 100}, 305  (1983).

\bibitem{concatenation}
A. Majumdar, E. P. Zuiderweg, {\it J. Magn. Reson. A} {\bf 11} 3, 19 (1995).

\bibitem{3-spin} N. Khaneja, S.J. Glaser, R.W. Brockett, 
{\it Phys. Rev. A} {\bf 65}, 032301 (2002).

\end{thebibliography}
\end{document}